\documentclass[prl, aps, twocolumn]{revtex4-1}
\usepackage{graphicx,graphics,psfrag,amsmath,calc,mathtools}
\usepackage{epsfig, bm}
\usepackage{color,comment}
\begin{document}
\date{\today}

\title{Crossovers and quantum phase transitions in two-band superfluids: \\
The evolution from BCS to Bose pairing by tuning interactions and band offset}
  
\author{Yue-Ran Shi$^{1,2}$}
\author{Wei Zhang$^{1}$} %\thanks{wzhangl@ruc.edu.cn}
\author{C. A. R. S\'a de Melo$^2$} % \thanks{carlos.sademelo@physics.gatech.edu}
\affiliation{$^1$Department of Physics, Renmin University of China, Beijing 100872, China}
%\affiliation{$^2$Beijing Academy of Quantum Information Sciences, Beijing 100193, China}
%\affiliation{$^3$Beijing Key Laboratory of Opto-electronic Functional Materials and
%Micro-nano Devices, Renmin University of China, Beijing 100872,China}
\affiliation{$^2$School of Physics, Georgia Institute of Technology, Atlanta, GA 30332, USA}
\date{\today}

\begin{abstract}
We show that in two-band $s$-wave superfluids it is possible to induce
quantum phase transitions (QPTs)
by tuning intraband and interband $s$-wave interactions, in sharp contrast to single-band
$s$-wave superfluids, where only a crossover between Bardeen-Cooper-Schrieffer (BCS)
and Bose-Einstein condensation (BEC) superfluidity occurs.
For non-zero interband and attractive intraband interactions, we demonstrate
that the ground state has always two interpenetrating superfluids possessing
three spectroscopically distinct regions
where pairing is qualitatively different: (I) BCS pairing in both bands (BCS-BCS),
(II) BCS pairing in one band and BEC pairing in the other (BCS-BEC),
and (III) Bose pairing in both
bands (BEC-BEC). Furthermore, we show that by fine tuning the interband interactions
to zero one can induce QPTs in the ground state between
three distinct superfluid phases. There are two phases where only one band is
superfluid ($S_1$ or $S_2$), and one phase where both bands are superfluid $(S_1 + S_2)$,
a situation which is absent in one-band $s$-wave systems. Lastly, we suggest
that these crossovers and QPTs may be observed in
$^{173}$Yb and $^{87}$Sr.
\end{abstract}

\maketitle

{\it Introduction:} The evolution from Bardeen-Cooper-Schrieffer (BCS) to Bose pairing
in one-band superfluids is a topic of intensive recent
experimental~\cite{moritz-2021, roati-2021, vale-2020, turlapov-2019},
and theoretical~\cite{brand-2021, ma-2021, bulgac-2020, castin-2019} research,
because it is the simplest system addressing the deep theoretical
connection between BCS
and Bose superfluidity~\cite{leggett-1980, nsr-1985, sademelo-1993}
that arises in many areas of physics:
condensed matter (superconductivity),
atomic physics (ultracold Fermi superfluids), astrophysics (superfluid neutron stars) and
quantum chromodynamics (color superconductivity)~\cite{sademelo-2008}.
Unfortunately, in quantum chromodynamics, astrophysics and condensed matter
the tunability of interactions is extremely limited or inexistent.
However, for low density one-band Fermi atoms ($^{6}$Li or  $^{40}$K)  it is possible
to tune $s$-wave interactions and study the crossover from BCS to Bose-Einstein
condensation (BEC) superfluidity~\cite{jin-2003, grimm-2003, hulet-2003, ketterle-2003},
where large Cooper pairs evolve into tightly bound pairs, when interactions
change from weak to strong.

Although the BCS-BEC crossover is interesting, its physics 
is not as striking as that occurring in quantum phase transitions (QPTs),
where singular behavior emerges.
In one-band superfluids, topological QPTs were theoretically predicted
as a function of interaction strength for higher angular momentum pairing,
such as $p$- or $d$-wave~\cite{duncan-2000, botelho-2005a, read-2000, botelho-2005b},
leading to superfluid phases in the BCS and BEC regimes
which are qualitatively different. However, the experimental observation of this phenomenon
in cold gases has failed, because $p$-wave fermion pairs dissociate by tunneling out of the
centrifugal barrier,
that is, their lifetime is just not long enough
to observe superfluidity~\cite{jin-2007,vale-2008, mukaiyama-2008}.
This experimental fact, makes it impossible to study
predicted QPTs in the superfluid state of $p$-wave or
higher angular momentum pairing~\cite{iskin-2006a} as a function of interactions.

In this paper, we propose an alternative idea to study elusive QPTs
between qualitatively different superfluid states
during the BCS to BEC evolution: tune only $s$-wave interactions, but enlarge the
Hilbert space of states to two bands.
The tuning of $s$-wave interactions creates stable fermion pairs, while the existence
of two bands allows for the emergence of QPTs. Experimental
candidates include systems with four internal states such as 
$^{6}$Li and $^{40}$K, where interactions may be tuned via
magnetic Feshbach resonances~\cite{jin-2003, grimm-2003, hulet-2003, ketterle-2003},
and $^{173}$Yb~\cite{takahashi-2007, julienne-2008} and
$^{87}$Sr~\cite{killian-2010, schreck-2011},
where interactions may be tuned via optical
Feshbach resonances~\cite{folling-2015, fallani-2015}. 
We investigate a Fermi gas with two parabolic bands per spin label
(four internal states) separated by a band offset
$\varepsilon_0$, whose physical origin can be a quadratic Zeeman shift.  We allow for
$s$-wave intraband and interband interactions, where the latter is described by
pair tunneling $J$.

We find two types of crossovers and two types of QPTs.
For non-zero $J$, the ground state phase diagram always exhibits
superfluidity in both bands for any chosen values of the intraband interactions,
this means that there are no QPTs, but there are two crossovers.
Typical crossover lines separate regions which are spectroscopically distinct
with respect to their quasiparticle excitation spectrum:
I) both bands have indirect gaps (BCS-BCS);
II) one band has an indirect gap and the other has a direct gap (BCS-BEC);
III) both bands have direct gaps (BEC-BEC).
The first type of QPT is a $0$-$\pi$ phase transition,
where the relative phases of the $s$-wave order parameters in the two bands
changes from $0$ $(J > 0)$ to $\pi$ $(J < 0)$.
However, the second type of QPT occurs
for $J = 0$, and leads to three different ground states as intraband interactions
are changed:
a)  two phases where only one band is superfluid ($S_1$ or $S_2$); and
b) one phase where both bands are superfluid $(S_1 + S_2)$.
Thus, QPTs in two-band $s$-wave superfluids
are found, rather than standard crossover physics in the evolution
from BCS to BEC superfluidity. 

\begin{figure}[t]
\centering{}
\includegraphics[width=0.9\columnwidth]{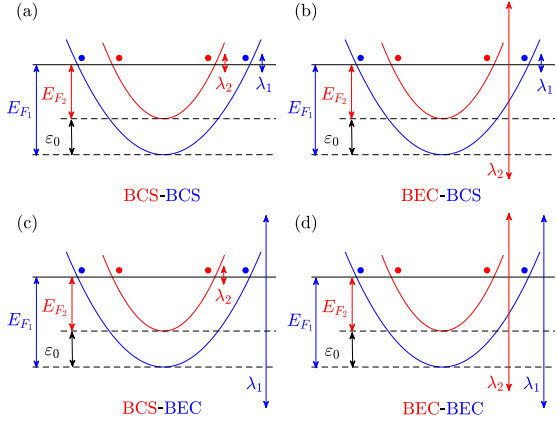}
\caption{
Energy dispersions and Fermi energies $E_{F_j}$
for $j = \{1, 2\}$: band 1 in blue is shifted down by  $\varepsilon_0$ with
respect to band 2 in red. Intraband interaction strengths are
$\lambda_j = \vert V_{jj} \vert  N_j$, where $N_j$ is the number of particles in band $j$.
When $\lambda_j \ll E_{F_j}$ pairing is BCS-like, and when 
$\lambda_j \gg E_{F_j}$, pairing in BEC-like in the $j^{\rm th}$ band.
\label{fig:one}
}
\end{figure}

{\it Hamiltonian:} To explore QPTs in two-band $s$-wave superfluids,
we start from the Hamiltonian
\begin{equation}
\label{eqn:hamiltonian}
H =
\sum_{j {\bf k} s}
\xi_j ({\bf k})c^\dagger_{j {\bf k} s}c_{j {\bf k} s}
+
\sum_{i j{\bf k} {\bf k}^\prime {\bf q}}
V_{ij}({\bf k}, {\bf k}^\prime)
b^\dagger_{i {\bf k} {\bf  q}} b_{j {\bf k}^\prime {\bf q}},
\end{equation}
where pair operators 
$b_{j {\bf k}{\bf  q}}
=
c_{j, -{\bf k} + {\bf q}/2, \downarrow} c_{j ,{\bf k} + {\bf q}/2, \uparrow}$
are defined in terms of fermion operators $c_{j {\bf k} s}$
with band index $j = \{1, 2 \}$, momentum ${\bf k}$ and spin labels
$s = \{ \uparrow, \downarrow \}$.
We work in three dimensions (3D) and choose units where $\hbar = k_B = 1$.
The term 
$
\xi_j ({\bf k})
=
\varepsilon_j ({\bf k})
-
\mu
$
is the kinetic energy for band $j$ with respect to the chemical potential $\mu$,
with
$
\varepsilon_j ({\bf k})
=
\varepsilon_{j 0}
+
\frac{{\bf  k}^2}{2m_j},
$
where  $m_j$ is the band mass.
%, so that in general the masses $m_1$ and $m_2$ can be different.
We choose
$\varepsilon_{10}= 0$ and $ \varepsilon_{20} = \varepsilon_0 > 0$, where
$\varepsilon_0$  is the energy offset between the two bands, as
shown in Fig.~\ref{fig:one}: the solid blue line (solid red line) represents band 1 (band 2), and 
$E_{F_1} = k_{F_1}^2/2m_1$ $(E_{F_2} = k_{F_2}^2/2m_2)$ is the Fermi energy
with Fermi momentum $k_{F_1}$ $(k_{F_2})$.
In Eq.~(\ref{eqn:hamiltonian}),
$V_{ij} ({\bf k}, {\bf k}^\prime)$ 
are intraband and interband interactions.
The intraband interactions are written in the separable form 
$
V_{jj}({\bf k},{\bf k}^\prime)
=
V_{jj}
\Gamma_j ({\bf k})
\Gamma_j ({\bf k}^\prime)
$
with
$
\Gamma_j ({\bf k})
=
\left[
 1 + {\bf k}^2/k_R^2
\right]^{-1/2},
$
%%
%
%where the momentum scale $k_R$ guarantees convergence
%at large momentum, that is, when $\vert {\bf k} \vert \gg k_R$.
with $k_R \sim R^{-1}$. Here, $R$ is the interaction range in real space.
The symmetry factor $\Gamma_{j} ({\bf k})$ in
$V_{11} ({\bf k}, {\bf k}^\prime)$ and
$V_{22} ({\bf k}, {\bf k}^\prime)$ represents $s$-wave pairing interactions.
The interband terms $V_{12} ({\bf k}, {\bf k}^\prime) = V_{12}$
and $V_{21} ({\bf k}, {\bf k}^\prime) = V_{21}$ 
%are taken to be independent of ${\bf k}$ and ${\bf k}^\prime$ and
are Josephson couplings, where $s$-wave pairs tunnel between bands, that is, 
%in momentum space and change their band index $j$.
%We take $V_{12} ({\bf k}, {\bf k}^\prime) = V_{12}$ and $V_{21} ({\bf k}, {\bf k}^\prime) = V_{21}$,
there is a momentum space proximity effect, where superfluidity in one band
can induce superfluidity in the other.

{\it Physical picture:} The parameters of the Hamiltonian in Eq.~(\ref{eqn:hamiltonian})
are: masses $m_1$ and $m_2$,
interactions $V_{11} = - \vert V_{11} \vert$, $V_{22} = - \vert V_{22} \vert$,
and $V_{12} = V_{21} = - J$,
band offset $\varepsilon_0$, and chemical potential $\mu$ fixing the total number of
particles $N = N_1 + N_2$.
Next, we set  $m_1 = m_2 = m$, but the arguments
presented are based on energetics and are also valid 
for $m_1 \ne m_2$~\cite{sademelo-2014}.
To compare interactions to Fermi energies $E_{F_1}, E_{F_2}$
we write the interaction energy scales
$\lambda_1 = \vert V_{11} \vert N_1$, $\lambda_2 = \vert V_{22} \vert N_2$, and
$\lambda_{J} =  J \sqrt{N_1 N_2}$.

In Fig.~\ref{fig:one}, Fermi energies
$E_{F_1}$ and $E_{F_2}$ are compared to interaction energies
$\lambda_1$ and $\lambda_2$. The Josephson energy
scale $\lambda_J$, not shown in the figure, is considered to the smallest of all.
A simple analysis of these energy scales leads to four general outcomes.
The first case is illustrated in panel (a), where the pairing
energy scales $\lambda_1 \ll E_{F_1}$
and $\lambda_2 \ll E_{F_2}$ leading to BCS pairing in both bands (BCS-BCS), and pair
sizes $\xi_{1} \gg k_{F_1}^{-1}$ and $\xi_{2} \gg k_{F_2}^{-1}$,
where $k_{F_j}$ is the Fermi momentum
associated with band $j$.
The second case is illustrated in panel (b), where 
$\lambda_1 \ll E_{F_1}$ and $\lambda_2 \gg E_{F_2}$
leading to BCS pairing in band 1 and BEC pairing in band 2 (BCS-BEC), and pair
sizes $\xi_{1} \gg k_{F_1}^{-1}$ and $\xi_{2} \ll k_{F_2}^{-1}$.
The third case is illustrated in panel (c), where
$\lambda_1 \gg E_{F_1}$ and $\lambda_2 \ll E_{F_2}$
leading to BEC pairing in band 1 and BCS pairing in band 2 (BEC-BCS), and pair
sizes $\xi_{1} \ll k_{F_1}^{-1}$ and $\xi_{2} \gg k_{F_2}^{-1}$.
The fourth case is illustrated in panel (d), where
$\lambda_1 \gg E_{F_1}$ and $\lambda_2 \gg E_{F_2}$
leading to BEC pairing in both bands (BEC-BEC), and pair
sizes $\xi_{1} \ll k_{F_1}^{-1}$ and $\xi_{2} \ll k_{F_2}^{-1}$.
The effect of $\lambda_J$ is to transfer
fermion pairs from one band to the other, thus guaranteeing that the
ground state is always superfluid with both bands
participating. Thus, when $\lambda_J \ne 0$, we can have only crossovers between
BCS-BCS, BCS-BEC, BEC-BCS and BEC-BEC regions.
The case of $\lambda_J = 0$ is very special, because it blocks pair transfer from one band
to the other, and allows for ground states where superfluidity exists not only
in both bands, but also in just one band, as either interactions or band offset are changed.
Thus, fine tuning $\lambda_J$ to zero allows for QPT's between different
superfluid phases rather than crossovers, even with only $s$-wave interactions.
As discussed below, this physical picture is further constrained by
$\mu$, because only
the total number operator
${\hat N} = \sum_{j {\bf k} s} c^\dagger_{j{\bf k}s} c_{j{\bf k}s} $ is conserved.

{\it Thermodynamic Potential:}
To obtain the thermodynamic potential
$\Omega = - T \ln {\cal Z}$, where ${\cal Z}$ is the grand canonical partition function,
we choose pairing to be independent of time and to occur at zero center-of-mass
momentum $({\bf q} = {\bf 0})$, that is, the pairing field is
$\Delta_j ({\bf q}) =  \Delta_{j 0} \delta_{{\bf q}{\bf 0}}$, where
$\Delta_{j0}$ is the order parameter for the $j^{\rm th}$ band.
This approximation leads to 
$
\Omega =
\Omega_{p}  +  \Omega_{c}.
$
The first term is
$
\Omega_{p} = 
-\sum_{ij} \Delta_{i0}^* g_{i j} \Delta_{j0}.
$
The second term, arising from the fermionic degrees of freedom, is
$
\Omega_{c} =
T\sum_{j {\bf k}}
\left\{
\beta
\left[
\xi_{j}({\bf k}) - E_{j}({\bf k})
\right]
- 2\ln \left[ 1+ e^{-\beta E_{j}({\bf k})} \right]
\right\},
$
where the quasiparticle excitation energy is
\begin{equation}
\label{eqn:quasi-particle-energies}
E_{j}({\bf k})= \sqrt{\xi^2_{j}({\bf  k}) + |\Delta_{j} ({\bf  k})|^2},
\end{equation}
with $\Delta_{j} ({\bf k}) = \Delta_{j 0}\Gamma_j ({\bf k})$.
When both $\vert \Delta_{10} \vert$ and $\vert \Delta_{20} \vert$ are non-zero,
$E_{j}({\bf k})$ is always gapped, but depending on $\mu$, this gap
can be indirect BCS-like, that is, at non-zero momentum; or direct BEC-like, that is, at
zero momentum. Spectroscopically, there are three regions:
(I) where $\mu > \varepsilon_0 $ and both $E_1 ({\bf k})$ and $E_2 ({\bf k})$ have
indirect BCS-like gaps (BCS-BCS); (II) where $\varepsilon_0 > \mu > 0$ and $E_1 ({\bf k})$
has an indirect BCS-like gap and $E_2 ({\bf k})$ has a direct BEC-like gap (BCS-BEC);
and (III) where $\mu < 0$ and both $E_1 ({\bf k})$ and $E_2 ({\bf k})$ have direct BEC-like gaps
(BEC-BEC).

Notice that, while $\Omega_{c}$ depends only on the moduli $\vert \Delta_{j0} \vert$,
we can write $\Omega_{p}$ in terms of the modulus
and phase of $\Delta_{j0} = \vert \Delta_{j0} \vert \exp \left( {i\varphi_j} \right)$
to obtain
$
\Omega_{p}
= - g_{11} \vert \Delta_{10} \vert^2 - g_{22} \vert \Delta_{20} \vert^2 
- 2 g_{12} \vert \Delta_{10} \vert \vert \Delta_{20} \vert \cos{\delta \varphi},
$
where $ \delta \varphi = \varphi_2 - \varphi_1$ is the relative phase between the
two order parameters. Here, $g_{11} = - V_{22}/ \det{\bf V}$,
$g_{22} = - V_{11}/ \det{\bf V}$, and  $g_{12} = - V_{12}/ \det {\bf V}$ with
$\det {\bf V} = (V_{11} V_{22} - V_{12}V_{21}) > 0$. Since 
$V_{12} = V_{21} = - J$ , then $g_{12} = J/ \det {\bf V}$ defines the sign of the prefactor
of $\cos {\delta \varphi}$. When $\vert \Delta_{10}\vert$ and $\vert \Delta_{20} \vert$
are non-zero and $J  > 0$ $(J < 0)$, the thermodynamic potential $\Omega$ is minimized when
the phases of the order parameters are the same (differ by $\pi$), that is,
$\varphi_2 = \varphi_1$ $(\varphi_2 = \varphi_1 \pm \pi)$. When $J = 0$,
$\varphi_1$ and $\varphi_2$ are completely independent. Thus, the limit $J \to 0$
is singular, that is, there is a phase transition between the $0$-phase with
$\delta \varphi = 0$ and the $\pi$-phase with
$\delta \varphi = \pm \pi$. This means that keeping
$V_{11}, V_{22}, \varepsilon_0,$ and $\mu$ fixed,
and switching $J \to -J$ leads to a $0$-$\pi$ QPT for any values
of the fixed parameters.

\begin{figure}[t]
\centering{}
\includegraphics[width=0.95\columnwidth]{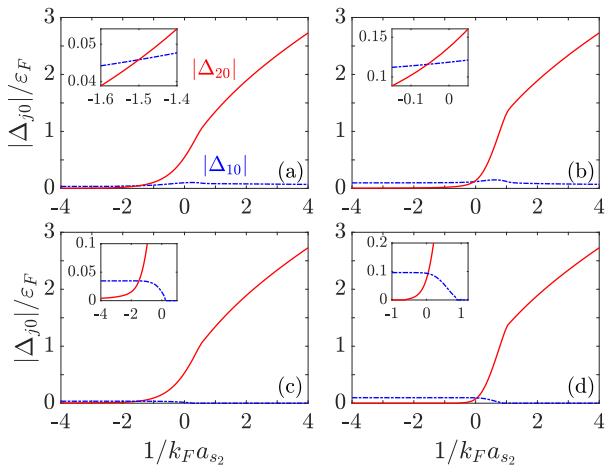}
\caption{
 (Color online) Order parameters
$\vert \Delta_{10}\vert/\varepsilon_F$ (dotdashed blue lines)
and $\vert \Delta_{20}\vert/\varepsilon_F$ (solid red lines) versus $1/k_Fa_{s_2}$
for fixed $1/k_Fa_{s_1} = -1.5$ and different values of
$(\varepsilon_0/ \varepsilon_F, J/\varepsilon_F)$:
(a) $(0, 10^{-3})$,
(b) $(0.9, 10^{-3})$,
(c) $(0, 0)$,
(d) $(0.9, 0)$.
}
\label{fig:two}
\end{figure}
\begin{figure}[t]
\centering{}
\includegraphics[width= 0.92\columnwidth]{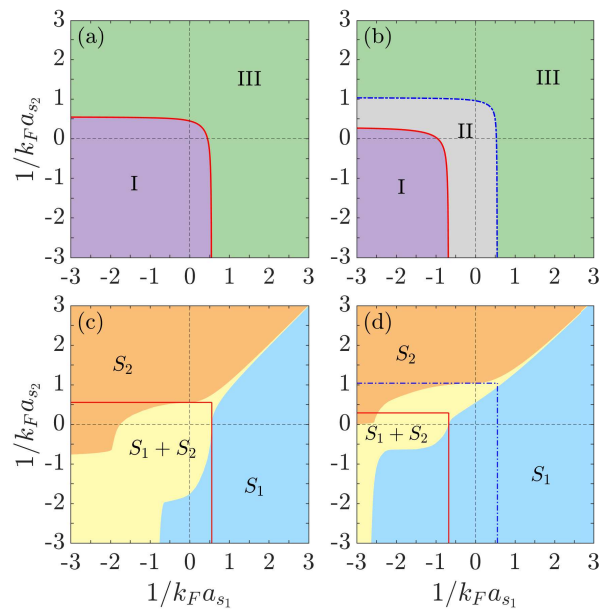}
\caption{
(Color online) Phase diagrams in $1/k_F a_{s_1}$ versus $1/k_F a_{s_2}$ plane
for different values of
$(\varepsilon_0/ \varepsilon_F, J/\varepsilon_F$):
(a) $(0, 10^{-3})$,
(b) $(0.9,  10^{-3})$,
(c) $(0, 0)$,
(d) $(0.9, 0)$.
In (a) and (b): the BCS-BCS region (I, purple), BCS-BEC region (II, gray),
and BEC-BEC region (III, green) are shown.
In (c) and (d):
three phases of $S_{1} + S_{2}$ (yellow) with  
$\vert \Delta_{10} \vert, \vert \Delta_{20} \vert \ne 0$, 
$S_1$ (blue) with $\vert \Delta_{10} \vert \ne 0$ and
$\vert \Delta_{20} \vert = 0$, and $S_2$ (orange) with $\vert \Delta_{10} \vert = 0$ and
$\vert \Delta_{20} \vert \ne 0$ are depicted. The solid red line is for $\mu = \varepsilon_0$ and
the dotdashed blue line is for $\mu = 0$.
}
\label{fig:three}
\end{figure}

{\it Order Parameters:} From the condition  $\delta \Omega/\delta \Delta_{i0}^* = 0$,
we obtain the order parameter equations 
\begin{equation}
\label{eqn:order-parameter-equation}
\Delta_{i0}
=
- \sum_{j{\bf k}}
V_{i j}\frac{ \Delta_{j 0} \vert \Gamma_j ({\bf  k}) \vert^2 }{2E_{j} ({\bf  k})}
\tanh \left[ \frac{\beta E_j ({\bf  k})}{2} \right]. 
\end{equation}
The number equation is
$N = - \partial \Omega /\partial \mu \vert_{T, V}$,
leading to
$
N = N_1 + N_2,
$
where
$
N_j
= 2 \sum_{{\bf k}} n_j ({\bf k})
$
is the number of particles in band $j = \{ 1, 2 \}$,
and
$
n_j ({\bf k}) =
\frac{1}{2}\left\{ 1 -  \frac{\xi_{j}({\bf k})}{E_{j}({\bf k})}
\tanh \left[ \frac{\beta E_{j}({\bf k})}{2} \right] \right\}
$
is the momentum distribution for each internal (spin) state of the $j^{\rm th}$ band.
For $J > 0$ with $\varphi_1 = \varphi_2$ or $J =0$ with $\varphi_1$ and $\varphi_2$ being
independent, we obtain $\vert \Delta_{j0}\vert$ and $\mu$
from the order parameter and number equations by writing $V_{11} = - \vert V_{11} \vert$
and $V_{22} = - \vert V_{22} \vert$ in terms of the $s$-wave scattering lengths
$a_{s_j}$~\cite{iskin-2005} via
$
\frac{1}{\vert V_{jj}\vert }
  =
-  \frac{m_j L^3}{4 \pi a_{s_j}}
+  \sum_{{\bf k}}
\frac{ |\Gamma_j({\bf k}) |^2 }{2 \varepsilon_j ({\bf k})}.
$
We use the total particle density $n = N/V$ to define an effective Fermi
momentum $k_F$ via $n = k_F^3/3\pi^2$ and an effective Fermi energy
$\varepsilon_F = k_F^2/2m$  as momentum and energy scales, since we
choose $m_1 = m_2 = m$ from now on. Note that $k_F^3 = k_{F_1}^3 + k_{F_2}^3$.
In Fig.~\ref{fig:two}, we show $\vert \Delta_{10} \vert/\varepsilon_F$ and
$\vert \Delta_{20} \vert / \varepsilon_F$ versus $1/k_F a_{s_2}$  for fixed
$1/k_F a_{s_1} = -1.5$, but different values of $(\varepsilon_0/\varepsilon_F, J/\varepsilon_F)$.
%: (a) $(0.0, 10^{-3}) $, (b) $ ( 0.9, 10^{-3}) $,
%(c) $(0.0, 0.0) $, (d) $(0.9, 0.0) $.
There are two messages from Fig.~\ref{fig:two}.
First, in panels (a) and (b), where $J/\varepsilon_F \ne 0$,
$\vert \Delta_{10} \vert/\varepsilon_F \ne 0$ and $\vert \Delta_{20} \vert / \varepsilon_F \ne 0$,
that is, both bands are always superfluid.
Second, in panels (c) and (d), where $J/\varepsilon_F = 0$, there are regions where
$\vert \Delta_{10} \vert/\varepsilon_F \ne 0$ and $\vert \Delta_{20} \vert / \varepsilon_F \ne 0$, 
that is, both bands are superfluid, but
when $1/k_{F}a_{s_2}$ is sufficiently large
$\vert \Delta_{10} \vert/\varepsilon_F = 0$ and $\vert \Delta_{20} \vert / \varepsilon_F \ne 0$,
that is, only band 2 is superfluid.

{\it Phase Diagrams:}
The ground state phase diagrams in the  $1/k_F a_{s_1}$
versus $1/k_F a_{s_2}$ plane, shown in Fig.~\ref{fig:three}, 
are determined by analyzing $\vert \Delta_{10} \vert$,
$\vert \Delta_{20} \vert$, and $\mu$.
The solid red (dotdashed blue) line corresponds to $\mu = \varepsilon_0$ $(\mu = 0)$.
In panels (a) and (b), where $J/\varepsilon_F \ne 0$, superfluidity arises in
both bands for all values of $1/k_F a_{s_1}$ and $1/k_F a_{s_2}$, that is,
$\vert \Delta_{10} \vert$ and $\vert \Delta_{20} \vert$ are always non-zero. Thus, there
are only crossovers between spectroscopically different superfluids phases:
(I) BCS-BCS (purple) with $\mu > \varepsilon_0$,
(II) BCS-BEC (gray) with $\varepsilon_0 > \mu > 0$ ,
and (III) BEC-BEC (green) with $\mu < 0$.
However, in panels (c) and (d), where $J/\varepsilon_F = 0$, there
are three different phases and QPTs between them.  The phases are
$S_1$ (blue) with $\vert \Delta_{10} \vert \ne 0$ and $ \vert \Delta_{20} \vert = 0$,
$S_2$ (orange) with $\vert \Delta_{10} \vert = 0$ and $ \vert \Delta_{20} \vert \ne 0$,
and
$S_1 + S_2$ (yellow)
with $\vert \Delta_{10} \vert \ne 0$ and $ \vert \Delta_{20} \vert \ne 0$.
For $J/\varepsilon_F = 0$, there is no superfluid proximity effect,
thus, the strongest-coupled band depletes the weakest-coupled band forcing
the order parameter of the latter to zero.
At the boundaries between $S_1 + S_2$ and $S_1$ $(S_2)$,
$\vert \Delta_{20} \vert$ $(\vert \Delta_{10} \vert)$ vanish and
the transitions are continuous. Furthermore,
when $\mu < \varepsilon_0$ $(\mu < 0$), 
$N_2$ $(N_1)$ also vanish where $\vert \Delta_{20} \vert = 0$ $(\vert \Delta_{10} \vert) = 0$.
These calculations confirm previous conjectures~\cite{sademelo-2014} and shine light
on earlier works that missed the full phase diagrams containing double crossovers and
QPTs~\cite{perali-2019, peeters-2012, babaev-2012, babaev-2011, iskin-2007, iskin-2006b}.

%
%%[hb]
\begin{figure}[t]
\centering{}
\includegraphics[width= 0.92\columnwidth]{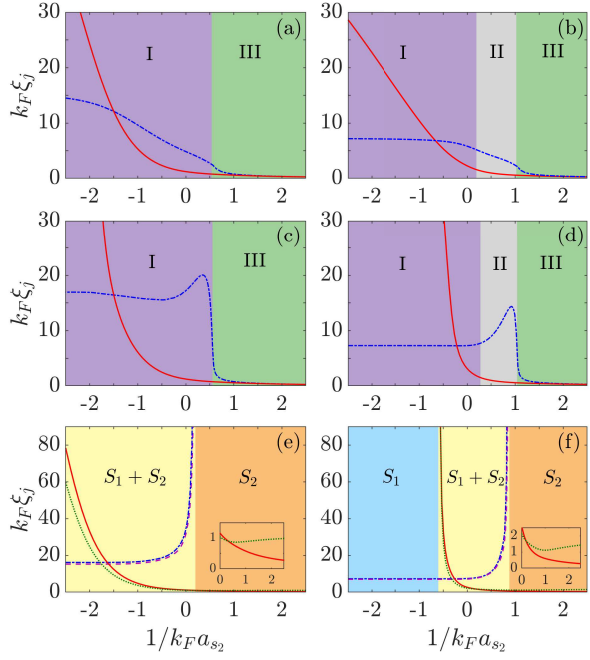}
\caption{
 (Color online) Pair sizes $k_F \xi_1$ (dotdashed blue line) and $k_F \xi_2$
(solid red line) versus  $1/k_F a_{s_2}$ 
for fixed $1/k_Fa_{s_1} = -1.5$ and different values of
$(\varepsilon_0/ \varepsilon_F, J/\varepsilon_F)$:
(a) $(0, 10^{-3})$,
(b) $(0.9, 10^{-3})$,
(c) $(0, 10^{-4})$,
(d) $(0.9, 10^{-4})$,
(e) $(0, 0)$,
(f) $(0.9, 0)$.
Background color codes are the same as in Fig.~\ref{fig:three}.
The value $k_F \xi_j =  1$ approximately separates the BCS-like ($k_F \xi_j \gg 1$)
and BEC-like $(k_F \xi_j \ll 1)$ regimes. The coherence lengths $k_F \xi_{1c}$ (dashed magenta line) and
$k_F \xi_{2c}$ (dotted green line) are shown in panels (e) and (f).
}
\label{fig:four}
\end{figure}

{\it Pair size:} To characterize further the spectroscopic regions $( {\rm I, II, III} )$
 and the QPTs for $J/\varepsilon_F = 0$, we discuss the pair sizes
$\xi_j$ within the $j^{\rm th}$ band~\cite{sademelo-1997, duncan-2000}:
\begin{equation}
 \xi_{j}^2 =
\left(
 \sum_{\bf k}
\phi_j ^* ({\bf k})  \left[ - \nabla_{\bf k}^2 \right] \phi_j({\bf k})
\right)
/
\sum_{\bf k} \vert  \phi_j ({\bf k}) \vert^2,
\end{equation}
where 
$
\phi_j ({\bf k}) =
\Delta_j ({\bf k}) / 2 E_j ({\bf k})
$
is the non-normalized pair wave function.
In Fig.~\ref{fig:four}, we show $k_F \xi_1$ (dotdashed blue line) and
$k_F \xi_2$ (solid red line) versus scattering parameter $1/k_F a_{s_2}$ 
for fixed $1/k_Fa_{s_1} = -1.5$ and different
values of $(\varepsilon_0/ \varepsilon_F, J/\varepsilon_F)$.
Panels (a) and (b) show that when $J/\varepsilon_F$ is sufficiently
large, that is, $J/\varepsilon_F = 10^{-3}$,  the pair sizes always decrease
as $1/k_F a_{s_2}$ increases. Thus, $k_F \xi_j$ monotonically
decreases from a
BCS-BCS region (I) to a BEC-BEC region (III) in
(a) and monotonically decreases from a BCS-BCS region (I)
to a BCS-BEC region (II) to a BEC-BEC region (III) in (b).
Panels (c) and (d) show that, for smaller $J/\varepsilon_F = 10^{-4}$,
$k_F \xi_1$ continues to decrease monotonically with $1/k_F a_{s_2}$,
however $k_F \xi_2$ first increases and then decreases before entering the
BEC-BEC region (III).
This non-monotonic behavior of $k_F \xi_2$ is simply a reflection of
the proximity to a QPT, where the order parameter
$\vert \Delta_{20} \vert$ is approaching zero.
The emergence of two QPTs is shown in panels (e) and (f), where
$J/\varepsilon_F = 0$. In this case, $k_F \xi_1$ ($k_F \xi_2$) increases
(decreases) monotonically with $1/k_F a_{s_2}$ and is zero
in the orange $S_2$ (blue $S_1$) region. The divergence in
$k_F \xi_j$ occurs as $\vert \Delta_{j0} \vert \to 0$.

{\it Ginzburg-Landau Theory:} As shown in Fig.~\ref{fig:two}, QPTs occur only
for $J/\varepsilon_F = 0$. In the vicinity of the phase boundaries between the
$S_1 + S_2$ (yellow) and $S_1$ (blue) or $S_2$ (orange) phases,
a Ginzburg-Landau (GL) theory is possible. Writing the order parameter as
$\Delta_j ({\bf q}) = \vert \Delta_{j 0}\vert \delta_{{\bf q},0}+ \Lambda_j ({\bf q})$,
and setting $\vert \Delta_{j0} \vert = 0$ at the appropriate boundary,
the GL thermodynamic potential becomes
%
%%
%\begin{equation}
$$
\Omega_{\rm GL} =
\Omega_{i}
+
\Omega_{j N} +
\int \frac{d^3 {\bf r}}{L^3}
\left[
\Lambda_j^* ({\bf r})
M_j ({\bf {\hat q}})
\Lambda_j  ({\bf r})  
+
b_j \vert \Lambda_j ({\bf r}) \vert^4
\right],
$$
%\end{equation}
%%
%
where $\Omega_{i}$ $(\Omega_{jN})$, with $i \ne j$, is the thermodynamic potential
of band $i$ $(j)$ which remains superconducting (becomes normal)
at the phase boundary. The fluctuation terms under
the integral are
$
M_j ({\bf {\hat q}}) =
a_j + c_j {\bf \hat q}^2/2m_j, 
$
and $b_j > 0$. The GL coherence
length $\xi_{j c}$ for pairing in the $j^{\rm th}$ band is 
$\xi_{j c}^2 = c_j/2m_j a_j$,
where $a_j = M_j ({\bf 0})$ and
$c_j = 2m_j \left[ \partial^2 M_j ({\bf q})/\partial {\bf q}^2 \right]_{{\bf q} = {\bf 0}}$,
with
$$
M_j ({\bf q})
= - g_{jj}
-
\sum_{{\bf k}, \lambda}
\vert \Gamma ({\bf k}) \vert^2
\alpha_j^{p\lambda}({\bf k}_{+}, {\bf k}_{-}) \beta_j^{p\lambda}({\bf k}_{+}, {\bf k}_{-}),
$$
where ${\bf k}_{\pm} = {\bf k} \pm {\bf q}/2$.
The index $\lambda = \{p, h \}$ represents quasiparticle $(p)$ or quasihole $(h)$ contributions.
The functions within the sum are 
$$
\alpha_j^{p\lambda} ({\bf k}_{+}, {\bf k}_{-})
=
\frac{
\tanh
\left[ E_{j} ({\bf k}_+)/2T \right]
\pm
\tanh \left[ E_{j} ({\bf k}_-)/2T \right]
}
{
E_{j} ({\bf k}_+)
\pm
E_{j} ({\bf k}_-)
},
$$
with the $+$ $(-)$  sign being for $\lambda = p$ $(\lambda = h)$, and 
$$
\beta_j^{p\lambda}({\bf k}_{+}, {\bf k}_{-})
=
\frac{1}{4}
\left[1 \pm \frac{\xi_j ({\bf k}_{+}) \xi_j ({\bf k}_{-})}{E_j ({\bf k}_{+}) E_j ({\bf k}_{-})}
\right]
$$
are the coherence factors.
%
%%
\begin{comment}
\textcolor{red}{Move expressions for $a_j$ and $c_j$ to a footnote, and also include
the expression for $b_j$ there. Then only give the final expression for $\xi_{jc}$ at
$T = 0$, and also at $T \ne 0$.}
\end{comment}
%%
%
Near the phase boundary at $T = 0$,
$\xi_{j c} = \xi_{j 0} \vert \eta_j - \eta_j^*  \vert^{-1/2}$,
where $\eta_j = 1/k_F a_{s_j}$ 
and $\eta_j^* = 1/k_F a_{s_j}^*$ is the critical interaction parameter.
When corresponding phase boundaries are crossed,  $\xi_{j c}$ diverges similarly to the pair size $\xi_{j}$,
signaling continuous phase transitions~\cite{footnote} over all phase boundaries
in the $1/k_F a_{s_1}$ versus $1/k_F a_{s_2}$ plane. This is illustrated 
in panels (e) and (f) of Fig.~\ref{fig:four}, where we can also see that in the BEC regime
$(1/k_F a_{s_2} \to \infty)$, the pair size $k_F \xi_{2} \to 0$, while the coherence length
$k_F \xi_{2c} \to C$, where $C \ne 0$.

{\it Conclusions:} We showed that, during the evolution from BCS to BEC superfluidity,
elusive quantum phase transitions (QPTs) occur by tuning $s$-wave interactions and
band offset in two-band superfluids. This in sharp contrast with single-band $s$-wave systems
where only a crossover is possible.
Our results may bypass long standing experimental difficulties in the search
of QPTs for ultracold fermions with one-band, where at least $p$-wave pairing
is required, but unfortunately $p$-wave Cooper pairs are short-lived.
In addition to QPTs, we have also established three spectroscopically distinct
superfluid regions - (I) BCS-BCS, (II) BCS-BEC, and (III) BEC-BEC -  
possessing crossovers between them, where pair sizes from each band
can be dramatically different. We analyzed pair sizes
and coherence lengths, within
the Ginzburg-Landau theory, and showed that they diverge at the
appropriate phase boundaries. Lastly, our results may motivate the experimental
search for multiband superfluidity and QPTs in ultracold $^{173}$Yb and $^{87}$Sr.

\acknowledgments
We thank the National Natural Science Foundation of China (Grants 11522436 $\&$ 11774425),
the Beijing Natural Science Foundation (Grant Z180013), and
 the National Key R$\&$D Program of China (Grant 2018YFA0306501) for financial support.


\begin{thebibliography}{2}

\bibitem{moritz-2021}
L. Sobirey, N. Luick, M. Bohlen, H. Biss, H. Moritz, and T. Lompe,
Observation of superfluidity in a strongly correlated two-dimensional Fermi gas,
Science {\bf 372}, 844 (2021).

\bibitem{roati-2021}
G. Del Pace, W. J. Kwon, M. Zaccanti, G. Roati, and F. Scazza, 
Tunneling transport of unitary fermions across the superfluid transition,
Phys. Rev. Lett. {\bf 126}, 055301 (2021).

\bibitem{vale-2020}
C. C. N. Kuhn, S. Hoinka, I. Herrera, P. Dyke, J. J. Kinnunen, G. M. Bruun,
and C. J. Vale,
High-frequency sound in a unitary Fermi gas,
Phys. Rev. Lett. {\bf 124}, 150401 (2020).

\bibitem{turlapov-2019}
M. Yu Kagan, and A. V. Turlapov,
BCS-BEC crossover, collective excitations, and hydrodynamics of superfluid
quantum liquids and gases,
Phys. -Usp. {\bf 62}, 215 (2019).

\bibitem{bulgac-2020}
A. Richie-Halford, J. E. Drut, and A. Bulgac,
Emergence of a Pseudogap in the BCS-BEC Crossover,
Phys. Rev. Lett. {\bf 125}, 060403 (2020).

\bibitem{ma-2021}
Hang Zhou, and Yongli Ma,
Thermal conductivity of an ultracold Fermi gas in the BCS-BEC crossover,
Scientific Reports, {\bf 11}, 1228 (2021).

\bibitem{brand-2021}
U. Ebling, A. Alavi, and J. Brand,
Signatures of the BCS-BEC crossover in the yrast spectra of Fermi quantum rings,
Phys. Rev. Research {\bf 3}, 023142 (2021).

\bibitem{castin-2019}
H. Kurkjian, S. N. Klimin, J. Tempere, and Y. Castin
Pair-breaking collective branch in BCS superconductors and superfluid Fermi gases,
Phys. Rev. Lett. {\bf 122}, 093403 (2019). 

\bibitem{leggett-1980} 
A. J. Leggett,
Cooper pairing in spin-polarized Fermi systems,
J. Phys. Colloq, {\bf 41}, 7 (1980).

\bibitem{nsr-1985}
P. Nozi\`eres, S. Schmitt-Rink,
Bose condensation in an attractive fermion gas: From weak to strong
coupling superconductivity,
J. Low Temp. Phys. {\bf 59}, 195 (1985).

\bibitem{sademelo-1993}
C. A. R. S\'a de Melo, M. Randeria, and J. R. Engelbrecht,
Crossover from BCS to Bose superconductivity: Transition temperature and
time-dependent Ginzburg-Landau theory,
Phys. Rev. Lett. {\bf 71}, 3202 (1993).

\bibitem{sademelo-2008}
C. A. R. S\'a de Melo,
When fermions become bosons: Pairing in ultracold gases
Phys. Today {\bf 61}, 45 (2008).

\bibitem{jin-2003}
M. Greiner, C. A. Regal, and D. S. Jin,
Emergence of a molecular Bose-Einstein condensate from a Fermi gas, 
Nature (London) {\bf 426}, 537 (2003).

\bibitem{grimm-2003}
S. Jochim, M. Bartenstein, A. Altmeyer, G. Hendl, S. Riedl, C. Chin,
J. Hecker Denschlag and R. Grimm,
Bose-Einstein Condensation of Molecules, 
Science {\bf 302}, 2101 (2003).

\bibitem{hulet-2003}
K. E. Strecker, G. B. Partridge, and R. G. Hulet,
Conversion of an atomic Fermi gas to a long-lived molecular Bose gas,
Phys. Rev. Lett. {\bf 91}, 080406 (2003).

\bibitem{ketterle-2003}
M. W. Zwierlein, C. A. Stan, C. H. Schunck, S. M. F. Raupach, S. Gupta, Z. Hadzibabic,
and W. Ketterle,
Observation of Bose-Einstein condensation of molecules,
Phys. Rev. Lett. {\bf 91}, 250401 (2003).

\bibitem{duncan-2000}
R. D. Duncan and C. A. R. S\'a de Melo,
Thermodynamic properties in the evolution from BCS to Bose-Einstein
condensation for a d-wave superconductor at low temperatures,
Phys. Rev. B {\bf 62}, 9675 (2000).

\bibitem{botelho-2005a}
S. S. Botelho and C. A. R. S\'a de Melo,
Lifshitz transition in d-wave superconductors,
Phys. Rev B {\bf 71}, 134507 (2005).

\bibitem{read-2000}
N. Read and D. Green,
Paired states of fermions in two dimensions with breaking of parity and
time-reversal symmetries and the fractional quantum Hall effect,
Phys. Rev. B {\bf 61}, 10267 (2000).

\bibitem{botelho-2005b}
S. S. Botelho and C. A. R. S\'a de Melo,
Quantum phase transition in the BCS-to-BEC evolution of p-wave Fermi gases,
J. Low Temp. Phys. {\bf 140}, 409 (2005).

\bibitem{jin-2007}
J. P. Gaebler, J. T. Stewart, J. L. Bohn, and D. S. Jin,
p-wave Feshbach molecules,
Phys. Rev. Lett. {\bf 98}, 200403 (2007).

\bibitem{vale-2008}
J. Fuchs, C. Ticknor, P. Dyke, G. Veeravalli, E. Kuhnle, W. Rowlands, P. Hannaford,
and C. J. Vale,
Binding energies of $^6$Li p-wave Feshbach molecules,
Phys. Rev. A {\bf 77}, 053616 (2008).

\bibitem{mukaiyama-2008}
Y. Inada, M. Horikoshi, S. Nakajima, M. Kuwata-Gonokami,
M. Ueda, and T. Mukaiyama,
Collisional properties of p-Wave Feshbach molecules,
Phys. Rev. Lett. {\bf 101}, 100401 (2008).
%Erratum Phys. Rev. Lett. 101, 139901 (2008)

\bibitem{iskin-2006a}
M. Iskin and C. A. R. S\'a de Melo,
Nonzero orbital angular momentum superfluidity in ultracold Fermi gases
Phys. Rev. A {\bf 74}, 013608 (2006).

\bibitem{takahashi-2007}
T. Fukuhara, Y. Takasu, M. Kumakura, Y. Takahashi,
Degenerate Fermi gases of Ytterbium
Phys. Rev. Lett. {\bf 98}, 030401 (2007).

\bibitem{julienne-2008}
M. Kitagawa, K. Enomoto, K. Kasa, Y. Takahashi, R. Ciuryło, P. Naidon, and P. S. Julienne,
Two-color photoassociation spectroscopy of ytterbium atoms and the precise
determinations of $s$-wave scattering lengths,
Physical Review A {\bf 77}, 012719 (2008).

\bibitem{killian-2010}
B. J. DeSalvo, M. Yan, P. G. Mickelson, Y. N. Martinez de Escobar, and T. C. Killian,
Degenerate Fermi gas of $^{87}$Sr,
Phys. Rev. Lett. {\bf 105}, 030402 (2010).

\bibitem{schreck-2011}
S. Stellmer, R. Grimm, and F. Schreck,
Detection and manipulation of nuclear spin states in fermionic strontium,
Phys. Rev. A {\bf 84}, 043611 (2011).

\bibitem{folling-2015}
M. H\"ofer, L. Riegger, F. Scazza, C. Hofrichter, D. R. Fernandes,
M. M. Parish, J. Levinsen, I. Bloch, and S. F\"olling,
Observation of an orbital interaction-induced Feshbach resonance in $^{173}$Yb,
Phys. Rev. Lett. {\bf 115}, 265302 (2015).

\bibitem{fallani-2015}
G. Pagano, M. Mancini, G. Cappellini, L. Livi, C. Sias, J. Catani, M. Inguscio,
and L. Fallani,
Strongly interacting gas of two-electron fermions at an orbital Feshbach resonance,
Phys. Rev. Lett. {\bf 115}, 265301 (2015).

\bibitem{sademelo-2014}
C. A. R. S\'a de Melo,
Erice Summer School, Multi-Condensates Superconductivity, p.27,
Superstripes Press, Rome, (2014).

\bibitem{iskin-2005}
M. Iskin and C. A. R. S\'a de Melo,
BCS-BEC crossover of collective excitations in two-band superfluids,
Phys. Rev. B {\bf 72}, 024512 (2005).

\bibitem{perali-2019}
Y. Yerin, H. Tajima, P. Pieri, and A. Perali, 
Coexistence of giant Cooper pairs with a bosonic condensate and
anomalous behavior of energy gaps in the BCS-BEC crossover
of a two-band superfluid Fermi gas,
Phys. Rev. B {\bf 100}, 104528  (2019).

\bibitem{peeters-2012}
L. Komendov\'a, Y. Chen, A. A. Shanenko, M. V.
Milo$\check{\text{s}}$evi\'c, and F. M. Peeters,
Two-band superconductors: Hidden
criticality deep in the superconducting state,
Phys. Rev. Lett. {\bf 108}, 207002 (2012).

\bibitem{iskin-2007}
M. Iskin and C. A. R. Sá de Melo,
Evolution of two-band superfluidity from weak to strong coupling,
J. Low Temp. Phys. {\bf 149}, 29 (2007).

\bibitem{iskin-2006b}
M. Iskin and C. A. R. Sá de Melo,
Two-band superfluidity from the BCS to the BEC limit,
Phys. Rev. B {\bf 74}, 144517 (2006).

\bibitem{babaev-2012}
M. Silaev and E. Babaev,
Microscopic derivation of two-component
Ginzburg-Landau model and conditions of its applicability in two-band systems,
Phys. Rev. B {\bf 85}, 134514 (2012).

\bibitem{babaev-2011}
M. Silaev and E. Babaev,
Microscopic theory of type-1.5 superconductivity
in multiband systems,
Phys. Rev. B {\bf 84}, 094515 (2011).

%
%%
\bibitem{sademelo-1997}
J. R. Engelbrecht, M. Randeria, and C. A. R. S\'a de Melo,
BCS to Bose crossover: Broken-symmetry state
Phys. Rev. B {\bf 55}, 15153 (1997)
%%
%

\bibitem{footnote}
The Ginzburg-Landau coherence length $\xi_{jc}$ and the pair size $\xi_j$ have different
physical meanings, the former is a measure of the phase coherence length of the superfluid
and the latter is a measure of the size of Cooper pairs.

\end{thebibliography}
\end{document}